\documentclass[aps,pra,preprint,showpacs,preprintnumbers,amsmath,amssymb,footinbib]{revtex4-1}
\usepackage{graphicx}
\usepackage{bm}
\usepackage[colorlinks=true, pdfstartview=FitV, linkcolor=red, citecolor=blue, urlcolor=blue]{hyperref}
\newcommand{\beq}{\begin{equation}}
\newcommand{\eeq}{\end{equation}}
\newcommand{\bea}{\begin{eqnarray}}
\newcommand{\eea}{\end{eqnarray}}

\begin{document}


\title{Collisional energy loss above the critical temperature in QCD}
\author{Shu Lin}
\email{slin@quark.phy.bnl.gov}
\affiliation{RIKEN/BNL Research Center, Brookhaven National Laboratory, 
Upton, NY 11973}
\author{Robert D. Pisarski}
\email{pisarski@bnl.gov}
\affiliation{
Department of Physics, Brookhaven National Laboratory, 
Upton, NY 11973}
\affiliation{RIKEN/BNL Research Center, Brookhaven National Laboratory, 
Upton, NY 11973}
\author{Vladimir V. Skokov}
\email{vladimir.skokov@wmich.edu}
\affiliation{
Department of Physics, Western Michigan University, 1903 W. Michigan 
Avenue, Kalamazoo, MI 49008
}
\begin{abstract}
We compute the collisional energy loss for a heavy quark
above the critical temperature in Quantum ChromoDynamics (QCD). 
We work in the semi Quark-Gluon Plasma, which assumes
that this region is dominated by the
non-trivial holonomy of the thermal Wilson line.
Relative to the result to leading order in perturbation theory,
at a fixed value of the coupling constant we 
generically we find that collisional energy loss is suppressed by 
powers of the Polyakov loop, $l<1$.  For small values of the loop, this 
suppression is linear for the scattering off of light quarks,
and quadratic for the scattering off of gluons, or for Compton scattering.
\end{abstract}
\maketitle

Experimentally the collisions of heavy ions at ultra-relativistic energies,
such as at the Relativistic Heavy Ion Collider (RHIC), or the Large Hadron
Collider (LHC), appear to be determined in large part by the behavior of
QCD at nonzero temperature.
At low temperatures, the confined phase 
can be modeled by a hadron resonance gas, while at high temperature,
a reasonable approach is to use a resummed perturbation theory.
However, for experiments both at the LHC, and especially at RHIC,
during most of the collision the temperatures probed are not far
about the transition temperature.  In QCD, this transition temperature
is that for the approximate restoration of chiral symmetry, $T_\chi$.

This intermediate region cannot be treated reliably either by a hadron
resonance gas, nor by (resummed) QCD perturbation theory.
One approach to this region is the 
``semi'' Quark-Gluon Plasma (QGP), where the ionization of color is
incomplete.  This region of partial deconfinement is modeled by
including a non-trivial holonomy for the thermal Wilson line,
by means of a matrix model
\cite{Pisarski:2000eq, *Dumitru:2003hp, *Diakonov:2004kc, *Diakonov:2003yy, *Diakonov:2003qb, *Dumitru:2004gd, *Oswald:2005vr, *Pisarski:2006hz, *Pisarski:2006yk, Meisinger:2001cq, *Meisinger:2001fi, Hidaka:2008dr, *Hidaka:2009hs, *Hidaka:2009xh, *Hidaka:2009ma, Dumitru:2010mj, *Dumitru:2012fw, Kashiwa:2012wa, *Pisarski:2012bj, *Lin:2013qu, *Bicudo:2013yza}.
A non-trivial holonomy implies that the expectation value of Polyakov loop
lies between its value in the confined phase, which is small (exactly zero
in the pure glue theory)
and that in the perturbative QGP, which is near one.
On a femtosphere one can show that this is
manifestly the appropriate effective theory
\cite{Ogilvie:2012is}.  In QCD, numerical simulations on the lattice
uniformly indicate that the Polyakov loop has such an intermediate
value between temperatures of $T_\chi$ and a few $T_\chi$
\cite{DeTar:2009ef, *Petreczky:2012rq}.

We note that there are other approaches to the semi-QGP.
These include 
quasi-particle models
\cite{Peshier:1995ty, *Peshier:2005pp, *Cassing:2007yg, *Cassing:2007nb, *Cassing:2009vt, *Castorina:2011ja, *Bratkovskaya:2011wp, *Ozvenchuk:2012fn},
which is indirectly related to a matrix model \cite{Kashiwa:2013gla}.
There are also Polyakov loop models, which take as variables not the
eigenvalues of the thermal Wilson line, but just its trace
\cite{Scavenius:2002ru, *Fukushima:2003fw, *Layek:2005fn, *Hell:2009by, 
*Fukushima:2010bq, *Buisseret:2011ms, *Horvatic:2010md, *Hell:2011ic}.
Other models includes gases of monopoles
\cite{Liao:2006ry, *Liao:2007mj, *Liao:2008jg, *Liao:2008vj, *Liao:2008dk, *Liao:2012tw, *Shuryak:2012aa}
and dyons
\cite{Diakonov:2004jn, *Diakonov:2005qa, *Diakonov:2007nv, *Diakonov:2009jq, *Diakonov:2010qg, *Fukui:2012gq, *Fukui:2012gp}.
There are also models involving bions, which are another type of matrix model
\cite{Unsal:2008ch, *Unsal:2007jx, *Shifman:2008ja, *Unsal:2008eg, *Shifman:2008cx, *Shifman:2009tp, *Poppitz:2009fm, *Simic:2010sv, *Unsal:2010qh, *Poppitz:2011wy, *Poppitz:2012sw}.

In this paper we directly use the results from numerical simulations on the
lattice to determine the eigenvalues of the Wilson line.  From this
we then compute the collisional energy loss for a heavy quark.
The computations are a straightforward
extension of those in ordinary perturbation theory.  

Our results have a simple physical interpretation.  A nontrivial
holonomy represents the fact that as the temperature decreases, the density
of particles with a given color charge decreases.  
This is obvious in the pure glue theory, where the probability to create
a particle with any color charge necessarily
vanishes in the confined phase.  That is, color is ``bleached''
in the confined phase.  With
dynamical quarks, at nonzero temperature there is always some small
probability to create particles with nonzero color charge.
Nevertheless, numerical simulations on the lattice indicate that this
probability is really rather small near the critical temperature, $T_\chi$.
In any case, particles in the adjoint representation, such as gluons,
are more strongly suppressed than quarks, 
which lie in the fundamental
representation.  To a good approximation for three colors, 
when the Polyakov loop
is small the density of gluons is proportional to the square of the
loop, while the density of quarks is proportional to a single power of
the loop.
 
We work to leading order in perturbation theory in a fixed background
field for the non-trivial holonomy.  For collisional energy loss, 
in the limit of a small value of the loop, 
we find that the scattering off of light quarks is suppressed by a single
factor of the loop, while that for gluons, or for Compton scattering in
a thermal bath, is quadratically suppressed.  
We expect that the suppression of scattering off of 
quarks and gluons near $T_\chi$
holds in any effective theory, although surely the details 
differ.  For the time being we defer a detailed comparison to experiment
to future study.

\section{Perturbative calculation of the collision energy loss with non-trivial holonomy}

\subsection{Introduction}
\label{intro_sec}

To represent nontrivial holonomy for a $SU(N_c)$ gauge group, 
we expand about a background, classical gluon field $A_0^{\rm cl}$, where
\begin{equation}
A_0^{\rm cl} = \frac{i}{g} \  {\rm diag}( Q^1, Q^2, \cdots,Q^{N_c}) \; .
\label{ansatz_Nc}
\end{equation}
Here $g$ is the coupling constant for the gauge field, so as
the background field is proportional to $1/g$,  
it is manifestly nonperturbative.
Further, the gluon field $A_0$ is not real, but purely imaginary.
We introduce such a mean field 
to model the effect of nontrivial holonomy, and so it 
should be understood as arising from an 
ensemble average over nonperturbative fluctuations.
Thus we do not attempt to derive from first
principles how this
field arises, but simply use results from lattice
simulations to determine  the $Q$'s.

Since the gauge group is $SU(N_c)$, the vector potential is traceless,
\begin{equation}
\sum_{a=1}^{N_c} {Q^a}=0.
\label{trless}
\end{equation}
The elements $Q^a$ are real, and we can assume that they are distributed
symmetrically about the origin.  (This is equivalent to assuming
the expectation value of the Polyakov loop, Eq. (\ref{trL}) below,
is real.  This is true if there is no net baryon density; otherwise
it is necessary to generalize the ansatz.) 

For three colors, as in QCD, this implies there is only one independent
variable, $Q$:
\begin{equation}
Q^a=(-Q,0,Q) \; .
\label{three_loop}
\end{equation}

The Wilson line in the temporal direction is
\begin{equation}
{   \mathbb{L}}(\vec{x})  
= {\cal P} \exp \left(  \int_0^{1/T} d \tau \; A_0(\vec{x},\tau)   \right)
\label{L} \; ;
\end{equation}
$T$ is the temperature, $\tau$ is the imaginary time, 
and $\cal P$ denotes time ordering.
The Wilson line is a unitary matrix, $\mathbb{L}^\dagger \mathbb{L} = 
{\bf{1}}$.  Under a gauge transformation $\Omega$, the Wilson
line transforms as ${   \mathbb{L}} \rightarrow
\Omega^\dagger \, {   \mathbb{L}} \, \Omega$.
The trace of the Wilson line is the Polyakov loop,
\begin{equation}
\ell(\vec{x}) = \frac{1}{N_c} \; {\rm tr}  \; \mathbb{L}(\vec{x}) \; ,
\label{trL}
\end{equation}
and is gauge invariant.  There are also higher loops, 
$(1/N_c) {\rm tr} \;\mathbb{L}^n$, which are obviously also
gauge invariant.  For a general field in $SU(N_c)$, there
are $N_c -1 $ independent loops.  

For three colors, under the mean field ansatz of Eq.
(\ref{three_loop}) there is one independent loop, which
we can take to be the simplest, 
\begin{equation}
\ell = \frac{1}{3} \left( 1 + 2  \cos\left( 
\frac{ Q}{T} \right) \right) \; .
\label{trLnc3}
\end{equation}

While the Wilson line is not gauge invariant, its eigenvalues are.
To leading order in weak coupling it suffices to deal with
the background $A_0^{\rm cl}$ field.  Beyond leading order it
is necessary to deal with the eigenvalues of the Wilson line,
which are gauge invariant.
Typically, lattice simulations do not measure the eigenvalues directly,
but only the Polyakov loop, which is a sum over the eigenvalues.  The
eigenvalues were measured directly in one recent study, 
\cite{Smith:2013msa}.  The results, however, agree with measurements
of the Polyakov loop.  

Physically the background field which generates non-trivial holonomy
can be thought of as an imaginary chemical potential for color.
The Bose-Einstein/Fermi-Dirac statistical distribution function for a gluons
and quarks are given, respectively, by 
\begin{eqnarray}
\label{df_g}
n^g_{ab}(E) &=& \frac{1}{\exp\left(|E-i Q^{ab}|_r/T\right)-1},\\ 
n^q_a(E) &=& \frac{1}{\exp\left(|E-i Q^{a}|_r/T\right)+1}  \; .
\label{df_q}
\end{eqnarray} 
Since quarks lie in the fundamental representation, their distribution
function involves only one color index, through $Q^a$.  For gluons in the
adjoint representation, a difference of two indices enters,
$Q^{a b} = Q^a - Q^b$.
We also introduce the notation 
$|z|_r = {\rm sign}({\rm Re} z)\  z$. 

These are the statistical distribution functions for emission into a 
thermal bath.
Those for absorption from a thermal bath are given by
\begin{equation}
\bar{n}_{ab}^{g} = 1 + n_{ab}^{g} 
\end{equation}
for gluons, and 
\begin{equation}
\bar{n}_a^{q} = 1 - n_a^{q} 
\end{equation}
for quarks.  The relative
minus sign is because quarks obey the Fermi-Dirac exclusion principle.

To illustrate the physics of non-trivial holonomy, consider the
sum over all colors for the quark statistical distribution function:
\begin{equation}
\sum_{a=1}^{N_c} n_a^q(E) = \sum_{a=1}^{N_c}
\frac1{1+\exp((E - i Q^a )/T)} = 
\sum_{n=1}^\infty (-)^{n+1} \; {\rm e}^{-n \, E/T} \; 
{\rm tr} \ { \mathbb{L}}^n \;.
\label{n}
\end{equation}

We first compute the number of quarks in the deconfined phase at
very high temperature.  At high temperature the theory is
essentially perturbative, and we can set all $Q$'s to vanish.  The
number of colored quarks is then
\begin{equation}
N_{\rm deconfined}^q 
= \int \frac{d^3 k}{(2 \pi)^3}  \;
\sum_{a=1}^{N_c} n^q_a(E) 
= N_c \left( \frac{3 \, \zeta(3)}{4 \, \pi} \right) \; ,
\label{Ndc}
\end{equation}
where the Riemann zeta-function
$\zeta(3) \approx 1.20206...$.  The numerical value of the right
hand side is not important, what we wish to emphasize is that 
as expected in the deconfined phase, the quark density 
is proportional to the number of quarks, $N_c$.

Contrast this with the background field in the confined phase
of the pure glue theory, where 
any Polyakov loop with nonzero $Z(N_c)$ charge vanishes.
The explicit $Q$ which produces the confined vacuum is
\begin{equation}
Q_{\rm conf}^a = \frac{\pi T}{N_c} \; (N_c + 1 - 2\, a) \;\;\; , \;\;\;
a = 1\ldots N_c \; .
\end{equation}
For three colors, Eq. (\ref{three_loop}), $Q = 2 \pi T/3$.
In the confined phase, all loops with nonzero $Z(N_c)$ charge vanish.
The only Polyakov loops which are nonzero are those which 
wrap around the imaginary time direction by an integral multiple of $N_c$.
These correspond to a type of ``baryon'':
\begin{equation}
\frac{1}{N_c} \; {\rm tr} \ \mathbb{L}^{k N_c}_{\rm conf}
=  (-)^{k (N_c+1) } \; .
\label{confined_loops}
\end{equation}
In the confined phase, then,
\begin{equation}
\sum_{a=1}^{N_c} n_a^q(E) = \frac{N_c}{1+\exp(N_c E/T)} \; .
\label{nc}
\end{equation}
For massless quarks, the energy is related to the momentum $k$ by $E = |k|$.
Integrating over the momenta, the total number of colored particles in
the confined phase is
\begin{equation}
N_{\rm confined}^q = 
\int \frac{d^3 k}{(2 \pi)^3} \; 
\sum_{a=1}^{N_c} 
n_a^q(E) = \frac{1}{N_c^2} \left( \frac{3 \, \zeta(3)}{4 \, \pi} \right) \; .
\label{Nc}
\end{equation} 
This computation illustrates several points.  In the limit of an infinite
number of colors, the number of quarks is $\sim N_c$ in the deconfined
phase, and very small, $\sim 1/N_c^2$, in the confined phase.  This ratio
is strictly zero only in the limit of infinite $N_c$.  For finite $N_c$,
there is a small density of quarks in the ``confined'' phase.  This matters
with dynamical quarks, where this density is nonzero.  This is the usual
observation that there is no strict order parameter for confinement in the
presence of dynamical quarks.  As mentioned previously, however, in practice
the density of quarks in the confined phase is small, at least as measured
by numerical simulations on the lattice for three colors and for two or
three flavors of quarks.

\subsection{Energy loss in the sQGP}

Consider a heavy quark of mass $M$ and energy $E$, where 
$E = \sqrt{\vec{p}\,^2 + M^2}$.  We assume that the heavy quark is moving
rapidly, with $p \gg M$.  
The energy loss per unit length $x$ is given by
\begin{equation}
\frac{dE}{dx}=\sum_i\frac{1}{2Ev}\int_k
\frac{n_i(k)}{2k}\int_{k'}\frac{{\bar n}_i(k')}{2k'}
\int_{p'} \; \frac{\omega }{2E' d} \;
|{\cal M}_i|^2 \; (2\pi)^4 \; \delta^{(4)}(P+K-P'-K')\   \; .
\label{dEdx}
\end{equation}
Here $P = (E,\vec{p}\,)$ is the four momentum of the incident heavy quark,
$P = (E',\vec{p}\,')$ that of the outgoing heavy quark, whilst
$K=(k,\vec{k}\,)$ and 
$K'=(k',\vec{k}')$ are the four momenta of the particles which the
heavy quark scatters off, and $\omega = E - E'$.
The integrals are over spatial momenta of the virtual particles,
\begin{equation}
\int_k = \int \frac{d^3 k}{(2 \pi)^3} \; ,
\end{equation}
and similarly for $\int_{k'}$ and $\int_{p'}$.
The index $i$ represents a label over the different types of particles,
including whether they are bosons or fermions/antifermions, as well as
color, flavor, and spin.
For the squared amplitude, $ {|{\cal M}_i|^2} $ in 
Eq.~\eqref{dEdx},
one sums over the initial and final spin states, and divides
by the degeneracy factor, $d=2$, of the incoming particle.

The computations of energy loss for a heavy quark in hot QCD were first
carried out by Braaten and Thoma \cite{Braaten:1991we}.
They showed that the integral over phase space can be simplified 
considerably; see also
Appendix A of Peign\'{e} and Peshier \cite{Peigne:2007sd}.
Using these simplifications,
\begin{equation}
\frac{dE}{dx}=\frac{1}{16\pi^2 \, E\, p}
\int_k\frac{n(k)}{2k}
\int^0_{t_{\rm min}} dt \; 
\int^{\omega_{\rm max}}_{\omega_{\rm min}} d\omega \; {\bar n}(k+\omega)\;
\frac{  \omega }{d \sqrt{\gamma}} \; \sum_i|{\cal M}_i|^2 \; ,
\label{dEdxsimp}
\end{equation}
Here $s$, $t$, and $u$ are the usual Mandelstam variables,
\begin{equation}
s = (P+K)^2 \;\;\; , \;\;\; t = (P - P')^2 
\;\;\; , \;\;\;  u = (P - K')^2 \; ,
\end{equation}
and we introduce the quantity $\gamma$,
\begin{equation}
\gamma = - \, \alpha^2 \, \omega^2 + \beta \, \omega + \delta    \; ,
\label{g}
\end{equation}
where
\begin{eqnarray}
\alpha &=& \frac{1}{p} \; (s - M^2) \; , \\ 
\beta &=& -\frac{2\, t}{p^2} \; 
\left[ E\, (s-M^2) - k\, (s+M^2)  \right] \; , \\
\delta &=& -\frac{t}{p^2} \; \left\{
t\left[ (E+k)^2 -s   \right] + 4 \, p^2 \, k^2 - (s-M^2-2\, E\, k)^2
\right\} \; .
\label{abc}
\end{eqnarray}

Since the square root of $\gamma$ enters into 
Eq. (\ref{dEdxsimp}), $\gamma$ must be positive.
Requiring that $\gamma > 0$ fixes the limit of 
integration over $t$ and $\omega$, with
$t: t_{\rm min} \rightarrow 0$, and 
$\omega: \omega_{\rm min} \rightarrow \omega_{\rm max}$.

In this article we only compute
the energy loss to leading logarithmic order. In this instance the 
distribution function ${\bar n}(k+\omega)$ can be replaced with $1$
\cite{Braaten:1991we,Peigne:2007sd}.  Thus we obtain
\begin{equation}
\frac{dE}{dx}=\frac{1}{16\pi^2 \, E\, p}
\int_k \; \frac{n(k)}{2k}
\int^0_{t_{\rm min}} dt \; 
\int^{\omega_{\rm max}}_{\omega_{\rm min}} d\omega \; 
\; \frac{\omega }{d \sqrt{\gamma}} \;  \sum_i|{\cal M}_i|^2 \; .
\label{dEdxsimp2}
\end{equation}

\subsection{Coulomb scattering}

\begin{figure*}[t]
\centerline{
\includegraphics[width=0.25\textwidth]{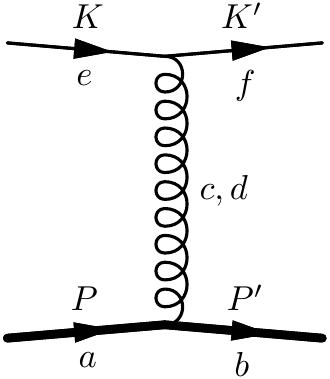} 
\hspace{0.2\textwidth}
\includegraphics[width=0.25\textwidth]{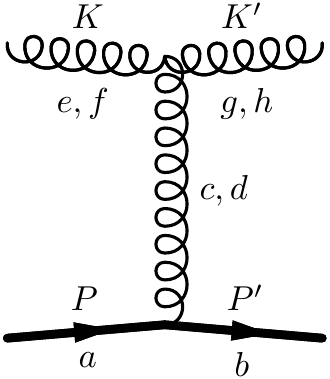}
}
\caption{
Scattering of a heavy quark (thick line) off of a light quark (left)
and a gluon (right) in the $t$-channel.  Historically, the diagram
on the left hand side is referred to as Coulomb scattering, 
while that on the right hand side is Compton scattering, off of a gluon,
in the $t$-channel.
}
\label{tch}
\end{figure*}

The amplitude for Coulomb scattering, of a heavy quark off of a light
quark in the thermal bath, is illustrated  in the left
hand side of Fig. (\ref{tch}).
This amplitude involves the color trace
\begin{equation}
\left( T^{cd}\right)_{ab} \left( T^{cd}\right)_{ef} 
\left( T^{d'c'}\right)_{ba}  \left( T^{d'c'}\right)_{fe} 
= \frac{N_c^2-1}{4 N_c}  \; .
\label{Ts}
\end{equation}
The summation in this expression must be performed with an open 
color index $e$, because
the energy loss depends on background
field through $Q^e$.  The amplitude reduces to 
\begin{equation}
\sum_i |{\cal M}_i|^2 
= \frac{8\, N_f \, g^4}{N_c} \left( \frac{N_c^2-1}{4 N_c} \right)
\; \left(\frac{2(s-M^2)^2+(u-M^2)^2+2M^2t}{t^2} \right) \; ,
\label{Couloubm}
\end{equation}
where $N_f$ is the number of light quark flavors, and thus
\begin{equation}
\left. \frac{dE}{dx} \right|_{Q}^{\rm qk}
= \frac{1}{16\pi^2 \, E\, p}
\sum_{e=1}^{N_c} \int_k \;
\frac{n^{\rm q}(k-i Q^e)}{2k}
\int^0_{t_{\rm min}} dt \; 
\int^{\omega_{\rm max}}_{\omega_{\rm min}} d\omega \;
\frac{ \omega }{d \sqrt{\gamma}} \; \sum_i|{\cal M}_i|^2 \; .
\label{dEdxsimpC}
\end{equation}
Here the subscript on $dE/dx$ refers to the dependence on the background
field $A_0^{cl}$ through $Q^e$.

The integration with respect to $\omega$ is
\begin{equation}
\int^{\omega_{\rm max}}_{\omega_{\rm min}} d\omega 
\; \frac{\omega}{\sqrt{\gamma}} = \pi \; \frac{\beta}{2 \, \alpha^3}\; .
\end{equation}
Integration with respect to the spatial momentum
$k$ is done by expanding the quark distribution function into a series
which starts with the Boltzmann term,
\begin{equation}
n^{\rm q}(k-i Q^e) 
= -\sum_{n=1}^\infty  (-)^n \; {\rm e}^{-n (k  + i Q^e)/T} \; .
\end{equation}

Keeping only the terms to leading logarithmic order, we find a very
simple result: the expression in the semi-QGP is an overall factor, which
depends upon the $Q$'s, times that for the perturbative QGP:
\begin{equation}
\left. \frac{dE}{dx} \right|_{Q}^{\rm qk}
= S^{\rm qk}(Q) \; 
\left. \frac{dE}{dx} \right|_{Q=0}^{\rm qk} \; ,
\label{dEdxCoulfinal}
\end{equation}
where the result in the perturbative QGP is
\begin{equation}
\left. \frac{dE}{dx} \right|_{Q=0}^{\rm qk}
= \alpha_{\rm s}^2 \; T^2 \; N_{f} \;
\frac{N_c^2-1}{12 N_c} \; \pi \; \ln\left(\frac{ET}{m_D^2} \right) \; ;
\label{dEdxpertq}
\end{equation}
$N_f$ is the number of light quark flavors.
We regulate the infrared logarithmic divergence 
of the integral over the Mandelstam variable $t$ by the gluon Debye 
mass.  

We note that in the semi-QGP that 
the gluon Debye mass depends upon the background field through the $Q$'s.
We can neglect this dependence because it only enters beyond leading
logarithmic order.

The $Q$-dependent factor in Eq.~\eqref{dEdxCoulfinal} is given by 
\begin{equation}
S^{\rm qk}(Q) = \frac{12}{\pi^2} \; \sum_{n=1}^{\infty} \;
\frac{(-)^{n+1}}{n^2} \; 
\left( \frac{  {\rm tr}\;  \mathbb{L}^n }{N_c}\right) \; .
\label{Sq}
\end{equation}
The superscript in $S^{\rm qk}(Q)$ denotes that it is
due to scattering off of light quarks.

In the perturbative regime this suppression factor equals unity, as
\begin{equation}
S^{\rm qk}(0)  = \frac{12}{\pi^2} \;
\sum_{n=1}^{\infty}  \frac{(-)^{n+1}}{n^2} = \frac {6}{\pi^2} 
\; \zeta(2) =  1 \; .
\label{SqPert}
\end{equation}  
In the confined phase of the pure glue theory, only
loops with $n = k N_c$ contribute, so that by
Eq. (\ref{confined_loops}), 
\begin{equation}
S^{\rm qk}(Q_{\rm conf})  = 
\frac{12}{\pi^2} \frac{1}{N_c^2 }\sum_{n=1}^{\infty}  
\frac{(-)^{n+1}}{n^2}  =    \frac{1}{N_c^2}. 
\label{SqConf}
\end{equation}
For physically relevant case, $N_c=3$,  and under the mean-field anzatz of Eq.~\eqref{three_loop}, 
by using the identity
\begin{equation}
\sum_{n=1}^{\infty} \frac{(-)^n}{n^2} 
\cos(2 \pi \, n\, x) =  \pi^2 \; \left( x^2- \; \frac{1}{12} \right)  \; 
\end{equation}
the suppression factor $S^{\rm qk}(Q)$ can be calculated analytically, 
\begin{equation}
S^{\rm qk}_{N_c=3}(Q = 2 \pi T q) 
=   \frac{4}{\pi^2} \; \sum_{n=1}^{\infty} \;
\frac{(-)^{n+1}}{n^2} \; 
\left( 1 + 2 \cos\left( 2 \pi n \, q \right)\right)
= 1 - 8 \, q^2 \; .
\label{sqNc3}
\end{equation}
Using Eq.~\eqref{trLnc3}, we have for the nearly  confining  and perturbative background field 
\begin{eqnarray}
&&S^{\rm qk}_{N_c=3}(Q \to Q_{\rm conf} ) =   \frac{1}{9} + \frac{8}{\pi \sqrt{3}} \ell+ {\cal O}(\ell^2)\;;\\
&&S^{\rm qk}_{N_c=3}(Q \to 0 ) =   1  - \frac{6}{\pi^2} (1-\ell) + {\cal O}((1-\ell)^2)\;,    
\label{sqNc3}
\end{eqnarray}
respectively. 

\subsection{Compton scattering}

\begin{figure*}[t]
\centerline{
\includegraphics[width=0.25\textwidth]{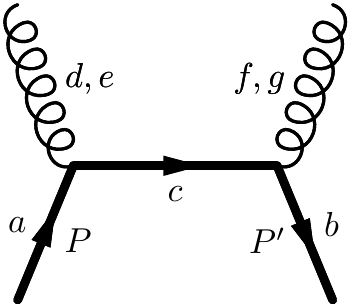}  
\hspace{0.2\textwidth}
\includegraphics[width=0.25\textwidth]{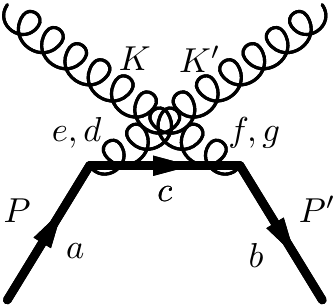}
}
\caption{
Compton scattering of a heavy quark (thick line) off of gluons
in the $s$-channel (left) and $u$-channel (right).  Only the diagram on the right hand side
generates a logarithm.
}
\label{uch}
\end{figure*}

There are three diagrams which contribute to what is termed
Compton scattering.
There is scattering off of a gluon in the $t$-channel, 
which is illustrated by the diagram
on the right hand side in Fig. (\ref{tch}).  There are also two
diagrams for scattering off of a gluon in the $s$-channel and $u$-channel, as 
illustrated in Fig. (\ref{uch}).  Among them, 
only the $t$-channel and $u$-channel generate leading logarithm contribution through small angles scattering. Furthermore, all cross terms between different channels do not lead to leading logarithm contribution, thus we will focus on the squared amplitude of $t$-channel and 
$u$-channel diagrams.

\subsubsection{Compton scattering in the $t$-channel}

For Compton scattering in the $t$-channel, the relevant diagram is
that on the right hand side of Fig. (\ref{tch}).  The color structure
which enters for this diagram is
\begin{equation}
\left( T^{cd}  \right)_{ab} f^{cd, ef,gh} 
\left( T^{d'c'}  \right)_{ba} f^{d'c', fe,hg}
= \frac{N_{\rm c}}{2} \left( 1 - 
\frac{1}{N_{\rm c}} \; \delta^{ef} \right) \; .
\label{colorTg}
\end{equation}
Again, there is no summation over the color indices $e$ and $f$, which
correspond to those for the gluon in the inital state, which the heavy
quark scatters off of.

The matrix element for scattering in the $t$-channel is
\begin{equation}
\sum_i |{\cal M}_i|^2 = -\frac{8 \;g^4}{N_{c}} 
\left( \frac{N_{\rm c}}{2} \right) \;
\left( 1 - \frac{1}{N_{\rm c}} \; \delta^{ef} \right) 
\left( \frac{(s-M^2)(u-M^2)}{t^2} \right) \; .
\label{Mgt}
\end{equation}

The expression for energy loss in the $t$-channel is
\begin{equation}
\left. \frac{dE}{dx} \right|_{Q}^{\rm gl}=
\frac{1}{16\pi^2\,E\, p}\sum_{e,f=1}^{N_c} \int_k \; 
\frac{n^{\rm g}(k-i [Q^e-Q^f])}{2k} \; 
\int^0_{t_{\rm min}} dt \; 
\int^{\omega_{\rm max}}_{\omega_{\rm min}} d\omega \;
\frac{\omega }{d \sqrt{\gamma}} 
\;  \sum_i|{\cal M}_i|^2 
\; .
\label{dEdxsimpCoul}
\end{equation}

Performing the integrals over $k$, $t$, and $\omega$ as before, again
the result is a $Q$-dependent factor times the perturbative result:
\begin{equation}
\left. \frac{dE}{dx} \right|_{\rm Q}^{{\rm gl, t-ch}}
= S^{\rm gl}(Q) \left. \frac{dE}{dx} \right|_{Q=0}^{\rm gl, t-ch} \; .
\label{dEdxtchan}
\end{equation}
The perturbative result is
\begin{equation}
\left. \frac{dE}{dx} \right|_{Q}^{\rm gl, t -ch}
= \alpha_s^2 \; T^2 \; (N_{\rm c}^2 -1) \; \frac{\pi}{6} \;
\ln \left(\frac{E\, T}{m_D^2} \right) \; .
\label{dEdxgtpert}
\end{equation}
The modification of the pertubative result in the semi-QGP is given
by a factor 
\begin{equation}
S^{\rm gl}(Q) = 
\frac{1}{N_{\rm c}^2-1} 
\left( \frac{6}{\pi^2}\sum_{n=1}^\infty  
\frac{|{\rm tr} \; \mathbb{L}^n|^2}{n^2} -1 \right)  \; .
\label{Sgt}
\end{equation}
In the perturbative QGP this factor is unity, 
\begin{equation}
S^{\rm gl}(0) = 
\frac{1}{N_{\rm c}^2-1} \left( \frac{6  }{\pi^2} 
\sum_{n=1}^\infty  \frac{N_c^2}{n^2} -1 \right) 
= \frac{1}{N_{\rm c}^2-1} \left( \frac{6  }{\pi^2} 
\zeta(2) N_c^2 -1 \right)= 1 \; ,
\label{SgtPert}
\end{equation}
as it must be.  The superscript in $S^{\rm gl}(Q)$
denotes that it is due to scattering off of a gluon.
We show in the next subsection
that to leading logarithmic order, the suppression factor
for scattering in the $u$-channel is the same as in the $t$-channel,
Eq. (\ref{dEdxuchan}).

In a confining background field, the suppression factor for
$t$-channel scattering is found to vanish,
\begin{equation}
S^{\rm gl}(Q_{\rm conf}) = 
\frac{1}{N_{\rm c}^2-1} \left( \frac{6  }{\pi^2} 
\sum_{n=1}^\infty  \frac{1}{n^2} -1 \right)= 0 \; .
\label{SgtConf}
\end{equation}

For three colors, $N_c=3$, using the anzatz Eq.~\eqref{three_loop} 
and the identity
\begin{equation}
\sum_{n=1}^{\infty} \; \frac{1}{n^2} \; \cos(2 \pi \, n \, x) 
= \frac{\pi^2}{12} \left(x^2-\, x+ \frac{1}{6}\right)  \; , 
\end{equation}
we obtain
\begin{equation}
S^{\rm gl}_{N_c=3}(Q = 2 \pi T q) 
= 1 - 3 \, q \left( 2 - 3 \, q  \right) \; .
\label{SgtNc3}
\end{equation}
In the limiting cases of a  confining  and perturbative background field we get 
\begin{eqnarray}
&&S^{\rm gl}_{N_c=3}(Q \to Q_{\rm conf} ) =  
\frac{27}{4} \; \ell^2(1-\ell) + {\cal O}(l^3) \;;\\
&&S^{\rm gl}_{N_c=3}(Q \to 0 ) =   1  - 
\frac{3 \sqrt{3}}{\pi} \; \sqrt{1-\ell}
+\frac{3\sqrt{3}}{4 \pi^2}\; (1-\ell) + {\cal O}((1-\ell)^2)\;.    
\label{sglNc3}
\end{eqnarray}

\subsubsection{Compton scattering in the $u$-channel}

The only diagram which generates a logarithm at leading order
is that on the right hand side of Fig. (\ref{uch}).  The color
structure for this diagram is
\begin{equation}
\left( T^{fg}  \right)_{ac} 
\left( T^{de}  \right)_{cb} 
\left( T^{fg}  \right)_{ac'} 
\left( T^{de}  \right)_{c'b} = 
\frac{C_f}{2} \;
\left( 1 - \frac{1}{N_{\rm c}} \delta^{de}\right)\; .
\label{TMgu}
\end{equation}
Here $C_f = (N_c^2-1)/(2 N_c)$ is the Casimir for the fundamental
representation.

The matrix element for scattering in the $u$-channel becomes
\begin{equation}
\sum_i |{\cal M}_i|^2 = 
-\frac{4 \; g^4}{N_{c}} \; \left( \frac{C_f}{2} \right)
\left( 1 - \frac{1}{N_{\rm c}} \delta^{de}\right)
\left(\frac{s-M^2}{u-M^2} \right) \; .
\label{Mgu}
\end{equation}

After integrating over $k$, $t$, and $\omega$, once again the result
is a $Q$-dependent factor times the perturbative result:
\begin{equation}
\left. \frac{dE}{dx} \right|_{\rm Q}^{\rm gl, u-ch}= S^{\rm gl}(Q) \left. 
\frac{dE}{dx} \right|_{Q=0}^{\rm gl, u-ch} \; .
\label{dEdxuchan}
\end{equation}
The result in the perturbative limit is
\begin{equation}
\left. \frac{dE}{dx} 
\right|_{Q=0}^{\rm gl, u-ch} = \alpha_s^2 \; T^2 \; C_f^2 \;
\frac{\pi}{6} \ln \left( \frac{ET}{M^2} \right) \; .
\label{dEdxgupert}
\end{equation}
We find that the color dependent factor in the semi-QGP is the same
in the $u$-channel as in the $t$-channel, given by Eq. (\ref{Sgt}).

\section{Complete Result}

\subsection{Extracting the loop from the lattice}

\begin{figure*}[t]
\centerline{
\includegraphics[width=0.48\textwidth]{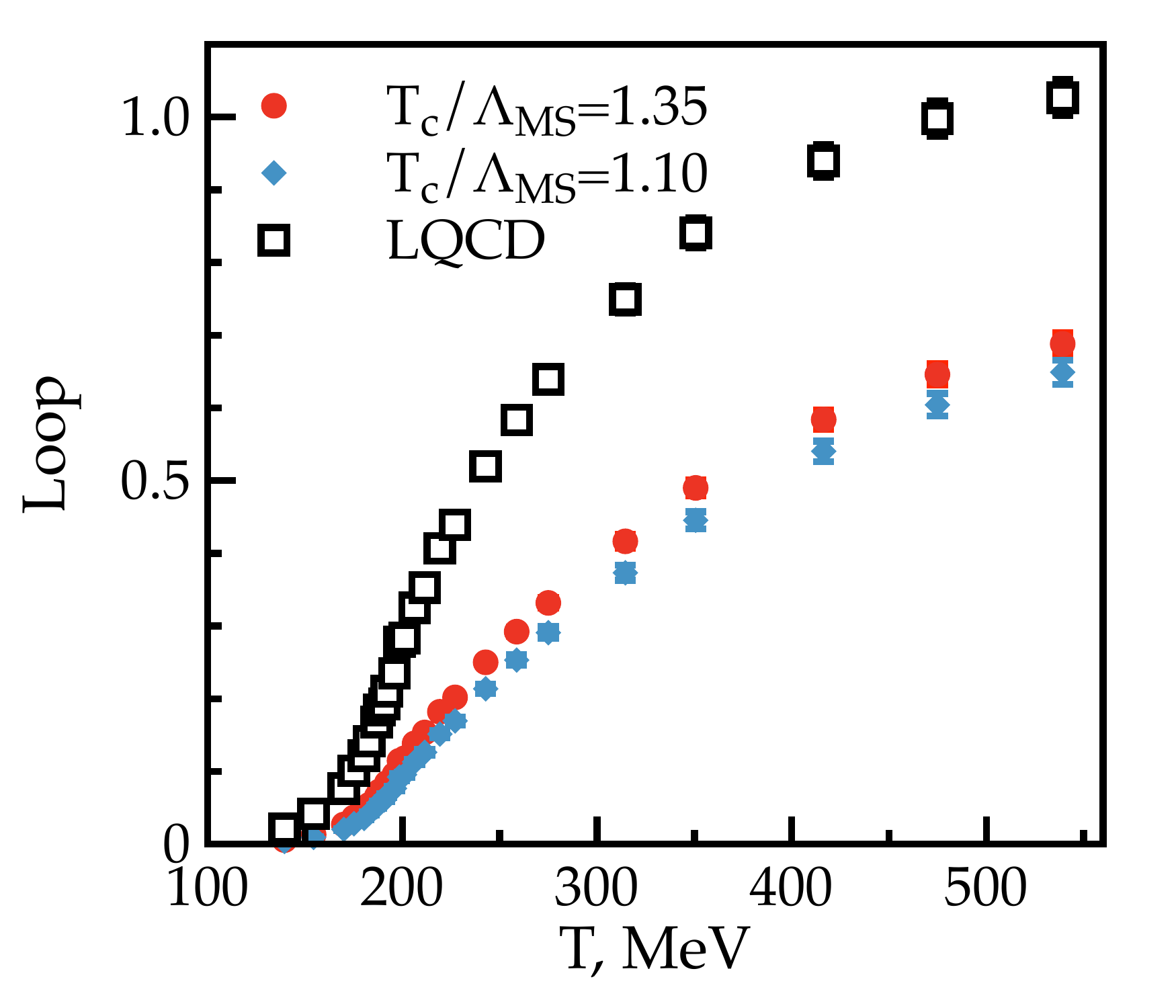} \hspace{0.02\textwidth}
\includegraphics[width=0.48\textwidth]{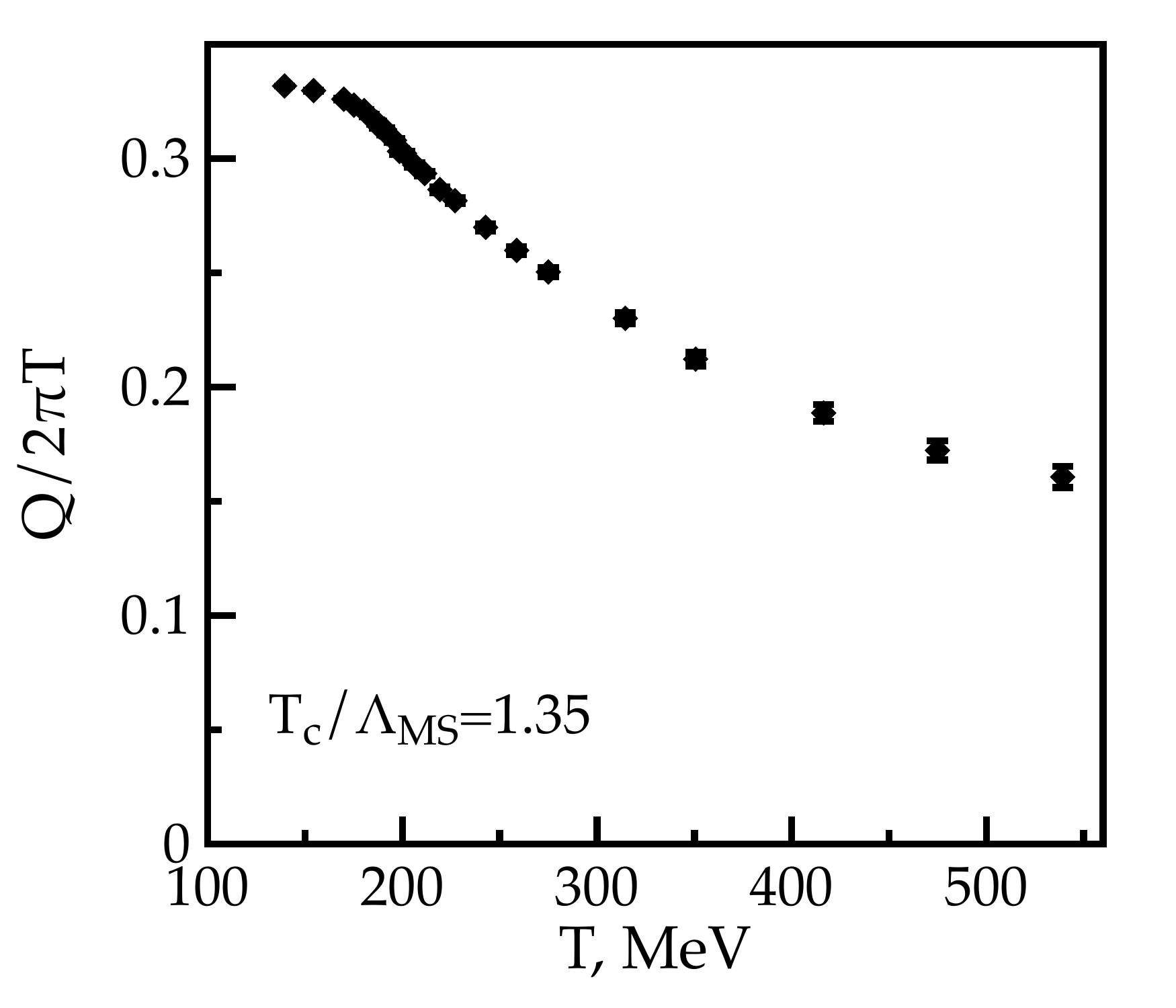}
}
\caption{
The left hand figure shows the
Polyakov loop from the lattice (LQCD), and its value
after removing perturbative corrections, as in Eq. (\ref{Lwpert}).
The result depends upon the value 
of the renormalization
mass scale, $\Lambda_{\overline{MS}}$.   The 
corresponding value of $Q$, Eq. (\ref{trLnc3}), is 
given in the figure on the right
hand side.
}
\label{Loopandq}
\end{figure*}

Besides non-perturbative contributions in the semi-QGP, the Polyakov loop
also receives contributions from ordinary perturbation theory,
\begin{equation}
\ell(Q=0) = \; 1 + \delta \, \ell(Q=0) \; .
\end{equation}
To order $\sim g^4$ \cite{Gava:1981qd,*Burnier:2009bk,*Brambilla:2010xn},
\begin{equation}
\delta \ell(Q=0) = + \; \frac{g^2 \, C_f \, m_E}{8\pi \, T} + 
\frac{g^4 C_f}{(4\pi)^2} 
\left[ - \; \frac{N_f}{2} \ln 2 + 
N_c \left( \ln \frac{m_E}{T}+\frac12 \right) \right] + {\cal O } (g^5) \; .
\label{Lpert}
\end{equation}
Notice that the leading contribution is positive.  This implies that
the expectation value of the loop exceeds unity.  While this cannot be
true classically, it occurs because of renormalization.

In Eq. (\ref{Lpert}), 
\begin{eqnarray}
g^2 &=&6 f_g \; , \\ 
m_E^2 &=& (2N_c + N_f)\  f_m\  T^2 \; ,
\label{g2}
\end{eqnarray}
and
\begin{equation}
f_{g,m} = \frac{4\pi^2} {(11 N_c - 2 N_f)
\left( \ln (4\pi T/\Lambda_{\overline{MS}}) 
- \gamma_{\rm E} + c_{g,m}\right)} \; ,
\label{fgm}
\end{equation}
where 
$\Lambda_{\overline{MS}}$ is the renormalization mass scale in
the modified minimal substraction scheme.  Lastly, the coefficients
for $c_g$ and $c_m$ are
\begin{eqnarray}
c_g &=& \frac{2N_f(4 \ln 2 -1) -11N_c}{2(11N_c-2N_f)},\\
c_m &=& \frac{4 N_f \ln 2} {11N_c -2N_f} 
- \frac{5N_c^2+N_f^2+9 \frac{N_f}{2Nc}}{(11N_c-2N_f)(2N_c+N_f)} \; .
\label{cm}
\end{eqnarray}

We assume that these perturbative corrections exponentiate,
\begin{equation}
\ell^{\rm total}(Q) = \exp \left [ \delta \ell(Q=0) \right] \; \ell(Q) \; .
\label{Lwpert}
\end{equation}
Even with $Q=0$, exponentiating the leading order corrections is
an assumption about those to higher order.  
Further, the corrections to $\sim g^3$ and $\sim g^4$ will certainly
change in the semi-QGP, when the $Q$'s are nonzero.  We do not
include this effect for the time being.

Thus we first compute $ \ell(Q) $ from Eq.~\eqref{Lwpert} and by using 
Eq.~\eqref{trLnc3} determine $Q(T)$. The results are shown in
Fig. (\ref{Loopandq}).

\subsection{Suppression factors in the semi-QGP}

Summing up the contributions from Coulomb scattering,
Eq. (\ref{dEdxCoulfinal}), 
from Compton scattering in the $t$-channel, Eq. (\ref{dEdxtchan}) 
and from Compton scattering in the $u$-channel, 
Eq. (\ref{dEdxuchan}), gives a total result for energy loss which is
\begin{eqnarray}
\frac{dE}{dx}  &=& \left( S^{\rm qk}(Q)   \; 
\alpha_s^2 \; T^2 \; \pi
\frac{N_f(N_c^2-1)}{12 \, N_c} \;
\ln\left(\frac{ET}{m_D^2}\right)\right.   \\
&+& \; S^{\rm gl}(Q)
\left. \left(  
   \frac{(N_{c}^2 -1)}{6} \ln \left(\frac{ET}{m_D^2}\right) + 
 \frac{C_f^2}{6} \ln \left(\frac{ET}{M^2}\right)
	\right) \right) 
\label{dEdxsum}
\end{eqnarray}

\begin{figure}[t]
\centerline{
\includegraphics[width=0.48\textwidth]{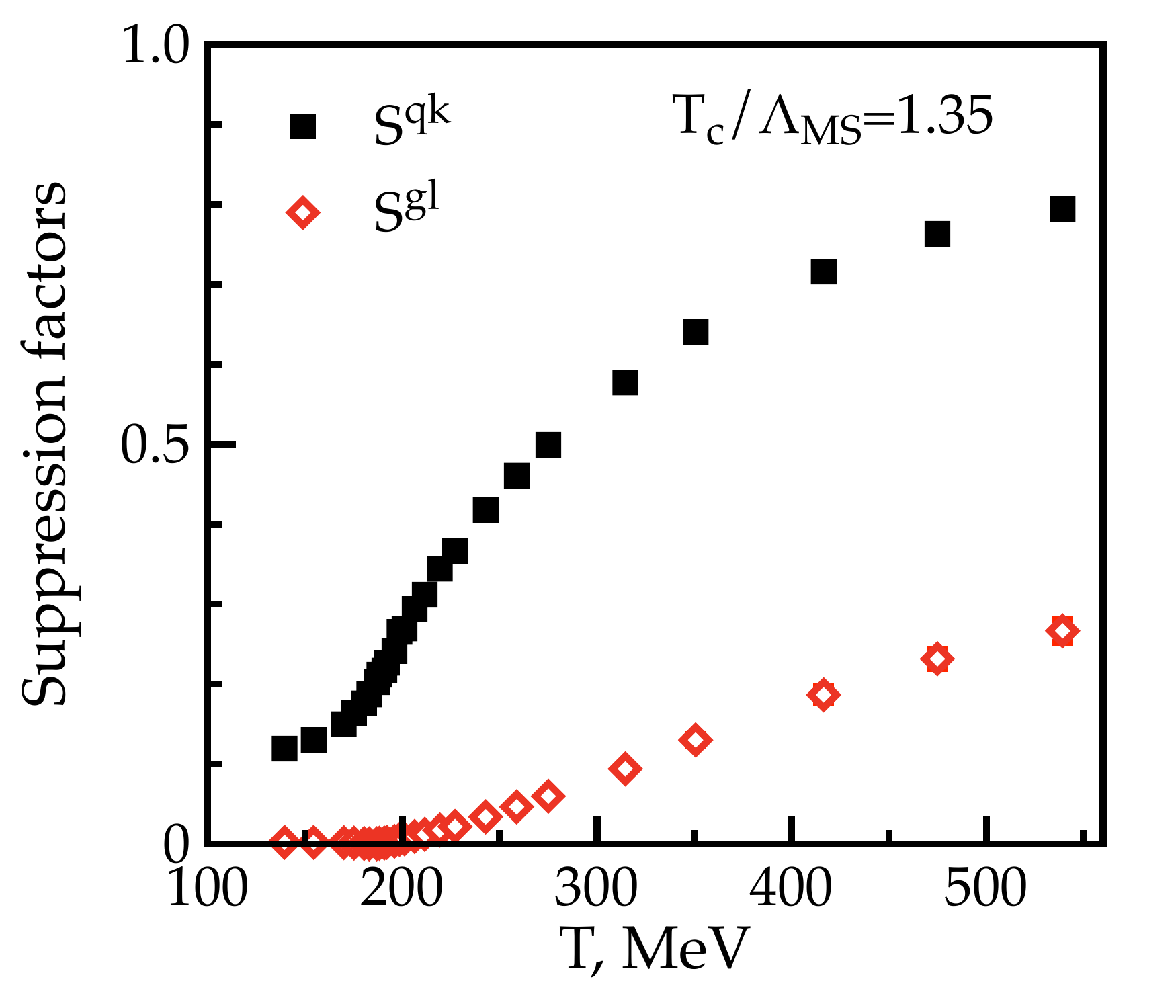}
}
\caption{
Suppression factors for quarks, $S^{\rm qk}$, and gluons, $S^{\rm gl}$, using
the values of the $Q$'s in Fig. (\ref{Loopandq}).  Notice that 
the suppression is much greater for gluons than for quarks.
}
\label{Sfactors}
\end{figure}

We can then use the results for the temperature dependence of $Q$ to
plot the suppression factors in the semi-QGP, versus the perturbative
results.  These are illustrated in Fig. (\ref{Sfactors}).  

For temperatures near $T_\chi$, where the expectation value of the Polyakov
loop is small, we find that the suppression of the gluon terms,
$S^{\rm gl}(Q)$, is much stronger than for the quark term,
$S^{\rm qk}(Q)$.  This is obvious from the corresponding expressions,
Eq. (\ref{Sgt}) for gluons, and Eq. (\ref{Sq}) for quarks.  For simplicity,
neglect corrections which are suppressed by factors of $1/N_c^2$ in
Eq. (\ref{Sgt}), since those are numerically small.  Then it is easy to
see that for small values of the loop, that
\begin{equation}
S^{\rm qk}(Q) \sim  \ell \;\;\; ; \;\;\;
S^{\rm gl}(Q) \sim  \ell^2 \; .
\end{equation}
Physically this is evident.  For small values of the loop, the density of
quarks is $\sim \ell$, while that of gluons is $\sim \ell^2$.  This is
simply because the quarks are in the fundamental representation, and the
gluons, in the adjoint.  In another way, in the double line notation
(which is useful at large $N_c$, but can be used at any $N_c$), quarks
have one line, and gluons, two lines.

We have only illustrated the suppression factors, and leave it for
later analysis to make a detailed comparison to experiment.  However,
our study shows that for temperatures which are probed at both RHIC and
even at the LHC, that the scattering off of light quarks completely dominates
over scattering off of gluons.  This is directly a manifestation of the
``bleaching'' of color in the semi-QGP, as the density of colored particles
decreases.  

\section{Conclusions and Outlook}

In this paper we computed the suppression of collisional energy loss in
the semi-QGP.  We find, on elementary and very general grounds, that 
scattering off of gluons is strongly suppressed in the semi-QGP, while
that of quarks is only moderately suppressed.

It is interesting that when the dust settles, we obtain rather simple
expressions for the collisional energy loss, as simple suppression factors
times the usual perturbative result.  This suggests that the same will
be true for other electromagnetic probes.  We have computed the 
effects of the semi-QGP upon both dilepton production, and on the
production of real photons, and will present these results shortly.  

These computations represent the first attempt to 
extend perturbative computations of quantities in thermal QCD 
to phenomenologically relevant temperatures of interest, building
crucially upon results from numerical simulations on the lattice.

\begin{acknowledgments}
S.L. is supported by the RIKEN Foreign Postdoctoral Researchers Program.
The research of R.D.P.  is supported
by the U.S. Department of Energy under contract \#DE-AC02-98CH10886.
We are indebted to Yoshimasa Hidaka for collaboration in the beginning of this
project and valuable comments. We also thank  Adrian Dumitru for discussions.
\end{acknowledgments}

%

\end{document}